\documentstyle[fleqn,epsf]{scr}
\begin{document}
\setcounter{page}{241}
\setlength{\textwidth}{41pc}
\title{Computing the real-time Green's Functions \\
of large Hamiltonian matrices$^{\dagger}$}
\author{Toshiaki Iitaka\\
Nanoelectronics Materials Group \\
Frontier Research Program, RIKEN \\
2-1 Hirosawa, Wako, Saitama 351-01, JAPAN 
}

\maketitle
\thispagestyle{headings}

\footnotetext[2]{Condensed from the article submitted to
Phys.\ Rev. E.}

\begin{center}
\bf
Abstract
\end{center}
A numerical method is developed for calculating the real time Green's
functions of very large sparse Hamiltonian matrices, which exploits the numerical solution of the inhomogeneous time-dependent Schr\"{o}dinger equation.
The method has a clear-cut structure reflecting the most naive
definition of the Green's functions,  and is very suitable to parallel and 
vector supercomputers.
The effectiveness of the method is illustrated by applying it to simple lattice models.

\section{Introduction}
\label{sec:intro}
In many fields of quantum physics, evaluation of the Green's functions constitutes the most important and difficult part of the theoretical treatment$^{1)}$.  
For example, to compute physical quantities of manybody systems in condensed matter physics, one should often calculate the Green's function of Hamiltonian matrices having a degree $N$ of $10^{6}$ or more.
Therefore efficient numerical algorithms, such as recursive Green's
function methods, quantum Monte Carlo methods, and the Lanczos methods 
have been developed and applied to various problems. 

Recursive Green's function methods$^{2)}$ have succeeded in evaluating dynamic quantities of relatively small systems by calculating directly the real-time Green's function.  For example, the conductance of quantum dots in chaotic, and regular regimes has been intensively investigated with these methods$^{3)}$. However, this scheme is prohibitive for huge Hamiltonian matrices because it requires computational time increasing rapidly as a function of the matrix size.

Quantum Monte Carlo methods$^{4,5)}$, which generate the imaginary-time Green's functions, have been successfully used for evaluating thermodynamic quantities of relatively large systems. For evaluating dynamic quantities such as conductivity, however, one has to rely on numerical analytic continuation (e.g., {\em maximum entropy method}$^{6)}$) from the imaginary-time Green's functions to the real-time ones. This procedure is, however, not unambiguous due to two reasons: one is the statistical errors originating from Monte Carlo sampling, which are amplified by numerical analytical continuation, and the other is the bias introduced by the default model in the maximum entropy method.

The Lanczos methods$^{7,8)}$ have been one of few reliable techniques
for evaluating dynamical responses of moderate-size Hamiltonian
matrices.The Lanczos methods use a linear transformation to a new
basis in which the Hamiltonian matrix has a tridiagonal form, and lead
to a continued fraction representation of the diagonal matrix elements
of the  Green's function. The drawback of these methods is the
numerical instability which may lead to {\it spurious
eigenstates}$^{9)}$. Recently, the Lanczos method has been extended to
the finite temperature case by introducing random sampling over the ground and excited states$^{10)}$.

In this paper, we present a new algorithm called {\em Particle Source Method} (PSM), which is based on the most naive and effective definition of the real-time Green's functions$^{12)}$. Namely, we calculate numerically the time-dependent Schr\"{o}dinger equation having a source term, and see how the wave function responds to the particle source.

This method has resemblances to the {\em Forced Oscillator Method}
(FOM), which has been developed by Williams and Maris$^{11)}$, and
applied to various classical and quantum problems$^{13,14)}$. The FOM
calculates the classical equations of motion of the coupled harmonic
oscillators driven by a periodic external force, where the matrix elements of the Hermitian matrix give the frequency and the coupling of the fictitious oscillators.
Our method is, however, much more clear-cut than their method, when
applied to quantum systems, since we calculate the time-dependent
Schr\"{o}dinger equation itself instead of the classical equations of
motion mapped from the quantum Hamiltonian matrix. The difference
between the two methods is analogous to the difference between the old quantum theories describing electronic states of an atom as an ensemble of fictitious harmonic oscillators and the modern quantum mechanics describing them by bra's and ket's.

Preceding to the present article, several authors have already solved
the homogeneous and inhomogeneous time-dependent Schr\"odinger equations
numerically $^{15,16,17,18,19)}$. Most of them are,
however, interested in launching wave packets in the computer and
watching them move around. Several of them tried to extract
time-independent quantities from the motion of the wave packets$^{18,19)}$. Unfortunately, their interest was limited in
obtaining several eigenvalues and eigenvectors of the Hamiltonian, but
not the Green's functions. As the result, they could obtain only the
exact peak position of the spectrum function (i.e., the imaginary part
of the diagonal elements of the Green's function), but could not calculate the correct shape of the spectrum function, the real part, and the off-diagonal elements. This is in contrast to our method, which can calculate both real and imaginary parts, and both diagonal and off-diagonal elements of the Green's function without calculating the eigenvalues and eigenvectors.

In section~\ref{sec:method}, we present the basic ideas of the PSM. In
section~\ref{sec:temperature}, we extend the PSM to the finite
temperature case. In Section~\ref{sec:examples}, we present numerical
examples to illustrate the effectiveness of the methods. A summary is given  in section~\ref{sec:summary}.

\section{Particle Source Method}
\label{sec:method}

\subsection{Single frequency calculation}
\label{subsec:single}
Let us introduce the time-dependent Schr\"{o}dinger equation with a time-dependent source term,
\begin{equation}
\label{eq:schroedinger.inhomogeneous.1}
i \frac{d}{dt}|\phi; t \rangle = H |\phi; t \rangle 
+ |j \rangle e^{-i(\omega+i\eta)t} \theta(t)
\end{equation}
where the wave function $|\phi; t \rangle$ and an arbitrary source $|j
\rangle$ are $N$-component complex vectors, the Hamiltonian $H$ is  an
$N \times N$ Hermitian matrix, $\omega$ is the frequency of the
source, and $\eta$ is a small positive imaginary part of the
frequency, which determines the resolution of frequency.
Note that this source term grows up exponentially as a function of time due to this small positive number, which simulates {\em adiabatic switching on} of the particle source. This adiabatic switching on, which has been absent in the preceding works$^{11,13,19)}$, is essential to calculate the exact shape of the Green's function as a function of energy.

The solution of this equation with the initial condition
$|\phi; t=0 \rangle = 0$ becomes$^{1)}$
\begin{eqnarray}
\lefteqn{|\phi; t \rangle }
\nonumber \\
&=& (-i) \int_0^{t} dt' e^{-iH(t-t') } |j \rangle e^{-i(\omega+i\eta)t'} \\
\label{green.1.1}
&=& \frac{1}{\omega+i\eta-H} \left( e^{-i(\omega+i\eta)t}-e^{-iHt} \right)|j \rangle \\
&\approx & \frac{1}{\omega+i\eta-H} e^{-i(\omega+i\eta)t} |j \rangle \\
\label{eq:green.1.2}
&=& G(\omega+i\eta) e^{-i(\omega+i\eta)t}|j \rangle 
\end{eqnarray}
where we have neglected the second term in the parentheses of (\ref{green.1.1}).
This approximation is justified by using sufficiently long time $t_1$ satisfying the condition
\begin{equation}
\label{eq:t1}
e^{-\eta t_1} < \delta
\end{equation}
where  $\delta $ is the required relative accuracy of the Green's function.

Then, from the Fourier transformation of (\ref{eq:green.1.2}), the Green's function operated on the ket $|j\rangle$ is obtained as
\begin{eqnarray}
\lefteqn{
\frac{1}{t_1} \int_0^{t_1} \!\!\! dt' \ | \phi; t \rangle  e^{i(\omega+i\eta)t}
}
\nonumber \\ 
&=& \frac{1}{t_1} \int_0^{t_1} dt'   G(\omega+i\eta ) |j \rangle  \nonumber \\
\label{eq:green.1.3}
&=& G(\omega+i\eta ) |j \rangle 
.
\end{eqnarray}
If only one or few matrix elements are necessary, we can calculate only these matrix elements as
\begin{eqnarray}
\lefteqn{
\frac{1}{t_1} \int_0^{t_1} dt' \langle i | \phi; t \rangle  e^{i(\omega+i\eta)t'}
} 
\nonumber \\
&=& \frac{1}{t_1} \int_0^{t_1} dt'  \langle i| G(\omega+i\eta ) |j \rangle  \nonumber \\
&=& \langle i| G(\omega+i\eta ) |j \rangle = G_{ij}(\omega+i\eta )
\end{eqnarray}
where $\langle i | $ is an arbitrary bra.

Since the numerical error due to the finite timestep is proportional to
$(\omega {\Delta t})^3$\ \ $^{17)}$, the best choice of $\omega$ is $\omega=0$.
The matrix elements with energy $\omega \ne 0$ can be obtained by calculating  the shifted Green's function at $\omega = 0$
\begin{equation}
G'(\omega=0;\eta)=\frac{1}{0+i\eta-H'}
\end{equation}
with the shifted Hamiltonian
\begin{equation}
H'=H-\omega I
\end{equation}
where $I$ is the unit matrix.

\subsection{Multiple frequency calculation}
Let us introduce the time-dependent Schr\"{o}dinger equation with a multiple frequency source term,
\begin{equation}
\label{eq:schroedinger.inhomogeneous.2}
i \frac{d}{dt}|\phi; t \rangle = H |\phi; t \rangle 
+ |j \rangle \left( \sum_{l=-L}^L e^{-i(\omega_l+i\eta)t} \right) \theta(t)
\end{equation}
where $\omega_l=l \Delta \omega$.

The solution of this equation with the initial condition
$|\phi; t=0 \rangle = 0$ becomes
\begin{eqnarray}
\lefteqn{|\phi; t \rangle}
\nonumber \\ 
&=& (-i) \int_0^{t} dt' e^{ -iH(t-t') } |j \rangle  \sum_l e^{-i(\omega_l+i\eta)t'}  \\
\label{eq:green.m1}
&=& \sum_l \frac{1}{\omega_l+i\eta-H} \left( e^{-i(\omega_l+i\eta)t}-e^{-iHt} \right) |j \rangle \\
&\approx & \sum_l \frac{1}{\omega_l+i\eta-H}|j \rangle e^{-i(\omega_l+i\eta)t} \\
\label{eq:green.m2}
&=& \sum_l G(\omega_l+i\eta ) |j \rangle e^{-i(\omega_l +i\eta)t}
\end{eqnarray}
where we have neglected the second term in the parentheses of (\ref{eq:green.m1}) as in the single frequency calculation. 
Then, from the Fourier transformation of (\ref{eq:green.m2}), the matrix elements of the Green's function are obtained as
\begin{eqnarray}
\lefteqn{
\frac{1}{t_2}\int_0^{t_2} dt' \langle i | \phi; t \rangle e^{i(\omega_{l'}+i\eta)t'}
} \nonumber \\
&=&\frac{1}{t_2}\int_0^{t_2} dt' \sum_l \langle i| G(\omega_l+i\eta ) |j
\rangle e^{-i(\omega_l-\omega_{l'})t'}  \\
\label{green.m3}
&=& 
\langle i| G(\omega_{l'}+i\eta ) |j \rangle 
\nonumber \\
&& + \sum_{ l\ne l'} \langle i| G(\omega_l+i\eta ) |j \rangle 
\frac{i\left
( e^{-i(\omega_l-\omega_{l'})t_2}-1\right)}{t_2(\omega_l-\omega_{l'})}
 \\
&\approx& G_{ij}(\omega_{l'}+i\eta )
\end{eqnarray}
where we have neglected the second term in (\ref{green.m3}).
This approximation is justified by using sufficiently long time $t_2$ satisfying the condition
\begin{equation}
\label{eq:t2}
 t_2 \Delta \omega > 1/\delta
.
\end{equation}

\subsection{Analysis of the Numerical Method}

\subsubsection{Solving the Schr\"odinger Equation}

To solve the time-dependent Schr\"odinger equation (\ref{eq:schroedinger.inhomogeneous.1}) numerically, we discretize it by using the leap frog method$^{15,16,17)}$,
\begin{eqnarray}
\label{eq:frog}
|\phi; t + {\Delta t} \rangle &=& -2 i {\Delta t} H |\phi; t \rangle + |\phi; t - {\Delta t} \rangle
\nonumber \\
&&-2 i {\Delta t} |j \rangle e^{-i(\omega+i\eta)t} \theta(t)
.
\end{eqnarray}
where ${\Delta t}$ is the time step. The time step is set as
\begin{equation}
{\Delta t} = \alpha /E_{max}
\end{equation} 
where $E_{max}$ is the absolute value of the extreme eigenvalue. We usually use the parameter $\alpha$ between $10^{-1}$ and $10^{-2}$.

Another method for the time-dependent Schr\"odinger equation is the Suzuki-Trotter decomposition of the time-evolution operator. Though the Suzuki-Trotter decomposition can be applied effectively only to a special class of Hamiltonian, it might have the advantage of the leap frog method. First it allows larger time step. Second it can be used with non-Hermitian Hamiltonian, such as the Hamiltonian with absorbing boundary condition.

\subsubsection{CPU time}

The computational time to calculate $G(\omega+i\eta)|j \rangle$ is estimated by the number $N_{prod}$ of matrix-vector products in (\ref{eq:frog}), which is equal to the integration time $t$ devided by time step ${\Delta t}$,
\begin{equation}
\label{eq:Nprod}
N^{prod}= \frac{t}{{\Delta t}}=\frac{ t E_{max} }{\alpha}
.
\end{equation}
Introducing (\ref{eq:t1}) into (\ref{eq:Nprod}), we obtain the number of matrix-vector products for the single frequency calculation
\begin{equation}
\label{eq:Nprod1}
N^{prod}_1= \frac{-\log \delta}{\alpha} \frac{E_{max}}{\eta}
.
\end{equation}
Therefore the relative error $\delta$ becomes exponentially small as a function of computational effort $N_{prod}$. On the other hand, the resolution $\eta$ is inversely proportional to $N_{prod}$, that is, we need longer CPU time for higher resolution.

Introducing (\ref{eq:t2}) into (\ref{eq:Nprod}), we obtain the number of matrix-vector products for the multiple frequency calculation
\begin{equation}
\label{eq:Nprod2}
N^{prod}_2= \frac{1}{\alpha\delta} \frac{E_{max}}{\Delta \omega}
.
\end{equation}
Therefore the relative error $\delta$ is inversely proportional to
$N_{prod}$, which means slower convergence than the single frequency
calculation. The distance between the frequencies to be measured,
$\Delta \omega$, is inversely proportional to $N_{prod}$, that is, we
need longer CPU time as we increase the number of the
frequencies. Actually, the integration time for the multiple frequency 
calculation should be the longer than $t_1$ and $t_2$. However,
because $t_1<t_2$ is usually satisfied, $t_2$ determines the CPU time
for the multiple frequency calculation.
Since the computational effort for a product of sparse matrix and vector is proportional to the matrix size $N$, the total computational time is estimated as
\begin{eqnarray}
T^{CPU}_1 &=& \frac{-\log \delta}{\alpha} \frac{E_{max}}{\eta} N N^{\omega} \\
T^{CPU}_2 &=& \frac{1}{\alpha\delta} \frac{E_{max}}{\Delta \omega} N
\end{eqnarray}
where $N^{\omega}$ is the number of the frequencies to be measured.

\subsubsection{Calculation of $G(\omega-i\eta)$}

So far, we have been calculating the Green's function whose frequency
has a positive imaginary part. When we need to calculate
$G(\omega-i\eta)$, we substitute $t$ by $-t$, and $\eta$ by $-\eta$ in
(\ref{eq:schroedinger.inhomogeneous.1}) and follow the same procedure
as described in section~\ref{subsec:single}. Then we obtain $G(\omega-i\eta)|j\rangle$.

\subsubsection{Product of the Green's functions}

Since $|j\rangle$ in (\ref{eq:schroedinger.inhomogeneous.1}) is an
arbitrary ket, we can repeat the calculation of the Green's function
by using a new source term,
\begin{eqnarray}
\lefteqn{
|j_2 \rangle e^{-i(\omega_2+i\eta_2)t}\theta(t)= 
}
\nonumber \\
&& A_1 G(\omega_1+i\eta_1) |j
\rangle e^{-i(\omega_2+i\eta_2)t} \theta(t)
\end{eqnarray}
where $A_1$ is an arbitrary operator whose matrix elements are known.
In general, we can calculate the matrix elements of a product
involving several Green's functions and other operators as
\begin{equation}
\label{eq:product}
\!\!\!\!\!\!\!\!\!\!\!\!\!\!\!
\langle i | A_n G(\omega_n \pm i\eta_n) \cdots 
A_2 G(\omega_2 \pm i\eta_2) 
A_1 G(\omega_1 \pm i\eta_1) 
 A_0 |j\rangle
.
\end{equation}

\subsubsection{Remote Eigenvalue Problem}

The remote eigenvalue problem pointed out in reference ${19)}$ does not appear in our methods, since we use very small time step in order to integrate the Schr\"odinger equation stably by using the leap frog method.

\subsection{Application to Manybody Problems}
\label{sec:manybody}

\subsubsection{Single-particle Green's function}
We can apply our methods for calculating the Green's functions of an
N-particle system at the ground state. As an example, let us see how
we can  calculate the retarded single-particle Green's function of an
N electron system on a finite lattice,
\begin{eqnarray}
\lefteqn{
{\cal G}_{ij}(\omega+i\eta) 
}
\nonumber \\
&=& (-i) \int_{-\infty}^\infty d\tau  
\langle g| \{a_i(\tau),a_j^\dagger(0)\} |g \rangle e^{i(\omega+i\eta )\tau}\theta(\tau)
\nonumber \\
\label{eq:green.manybody}
&=& + \langle g| a_i G(E_g+\omega+i\eta) a_j^\dagger |g \rangle 
\nonumber \\
&& - \langle g| a_j^\dagger  G(E_g-\omega-i\eta) a_i  |g \rangle 
\end{eqnarray}
where $a_i$ and $a_j^\dagger$ are the annihilation operator at site
$i$ and the creation operator at site $j$; $|g \rangle$ and $E_g$ are
the groundstate of the N electron system and its energy, respectively.
Since each term of (\ref{eq:green.manybody}) has the form of (\ref{eq:product}), we can calculate
${\cal G}(\omega)$ as follows:

First, we calculate the ground state $|g\rangle_N$ of the N-electron
system by using one of existing methods such as the Lanczos method,
the quantum Mote Carlo method, or the finite difference method$^{21)}$. 
The advantage of using the finite difference method is that it can
recycle most of the subroutine resource written for calculating the
Green's function since both programs solve the time-dependent
Schr\"odinger equation in the same way.

Second, we operate $a^\dagger_j$ to the ground state to obtain an
$N+1$ electron state, $|j\rangle_{N+1}= a^\dagger_j |g\rangle_N $. In
a similar way, we calculate another $N+1$ electron state, 
$|i\rangle_{N+1}= a^\dagger_i |g\rangle_N$.

Finally, the retarded Green's function is calculated in $N+1$-electron
subspace using the method in the previous subsection together with the state vectors $|j\rangle_{N+1}$ and $|i\rangle_{N+1}$.

\subsubsection{Optical Conductance}

The optical conductivity is expressed within the linear response theory 
as 
\begin{eqnarray}
\lefteqn{
\sigma_{xx}(\omega+i\eta) 
}
\nonumber \\
&=&
\frac{1}{\omega} \int_{-\infty}^\infty \!\!\!\! dt \  e^{i(\omega+i\eta) t} 
\langle g | j_x(t)j_x(0) |g \rangle 
\nonumber \\
&=& \frac{-2}{\omega} \ {\rm Im \ } \ \langle g|j_x G(\omega+E_g+i\eta)j_x |g \rangle
,
\end{eqnarray}
which we can calculate by using the PSM.
\section{Monte Carlo Particle Source Method}
\label{sec:temperature}
In this section, we extend PSM to finite
temperature case by using {\em Monte Carlo Particle Source Method} (MCPSM), a combination
of PSM and the Monte Carlo method for calculating the trace of a large matrix.
It turns out that, for sufficiently large systems, only a single configuration of
random variables suffices to evaluate the desired expectation value at 
finite temperatures.

\subsection{Monte Carlo Calculation of Trace}
Computing trace of a large matrix $A$ requires evaluation of $N$
diagonal elements of the matrix. Therefore it would take formidable
computational time if we try to use PSM for calculating the trace of
the product of operators including the Green's functions. A Monte Carlo
method to estimate trace of large matrices$^{11)}$ makes it
possible to evaluate the trace of this kind.

Let us introduce a set of random variables $\phi_n, (n=1,\cdots,N)$ that satisfy the relation 
\begin{equation}
\left\langle \left\langle \   e^{-i\phi_{n'}} e^{i\phi_{n}} \  \right\rangle \right\rangle \  = \delta_{n'n}
\end{equation}
where $\left\langle \left\langle \   \cdot \  \right\rangle \right\rangle \ $ implies statistical average. Then we define a random ket as
\begin{equation}
|\Phi \rangle = \sum_{n=1}^N |n\rangle e^{i\phi_n} 
\end{equation}
with the chosen basis set $\{ |n\rangle \}$.
Then the statistical average of $\langle \Phi | A | \Phi \rangle$ gives an approximation of the trace,
\begin{eqnarray}
\label{eq:trace.monte}
\lefteqn{
\left\langle \left\langle \   \langle \Phi | A | \Phi \rangle \  \right\rangle \right\rangle \  
}
\nonumber \\
&=&
\sum_n \langle n|A|n \rangle   
\nonumber \\
&&
+\sum_{n \ne n'} \left\langle \left\langle \   e^{i(\phi_n-\phi_{n'})} \  \right\rangle \right\rangle \  \langle n'|A|n \rangle \\
&=& {\rm tr}\left[ A \right]
.
\end{eqnarray}
The second term in (\ref{eq:trace.monte}) gives the statistical error
when the average is evaluated by the Monte Carlo method.
Assuming all non-zero matrix elements have the value of oder of $1$,
the first term in (\ref{eq:trace.monte}) becomes oder of $N$,
while the fluctuation of the second term becomes oder of
$\sqrt{N}$ for sparse matrices since the number of non-zero matrix
elements is oder of $N$ for sparse matrices. Therefore, the
statistical error of the trace will become small as $1/\sqrt{N}$.
For example, the statistical error becomes $10^{-3}$ for $N=10^{6}$,
which can be considered as small enough. If the statistical error with 
a single set of random variables is
not small enough, we can repeat the calculation with $M$ sets of
random variables $\phi^{(m)}_n, (n=1,\cdots,N)$ where $(m=1,\cdots,M)$
and obtain the statistical error of order of $1/\sqrt{MN}$

If the operator $A$ is Hermitian,  the imaginary part of the
statistical error becomes zero during the Monte Carlo process since
\begin{eqnarray}
\lefteqn{
\sum_{n \ne n'}  e^{i(\phi_n-\phi_{n'})}  \langle n'|A|n \rangle 
}
\nonumber \\
&=& \sum_{n > n'} 2 {\rm Re} \left(  e^{i(\phi_n-\phi_{n'})}  \langle n'|A|n\rangle \right)
.
\end{eqnarray}

This algorithm can be applied for evaluating, for example, the density of state,
\begin{eqnarray}
\rho(\omega) &=& \frac{-1}{\pi} \sum_n {\rm Im \ } G_{nn}(\omega + i\eta) \\
&=& \frac{-1}{\pi} {\rm Im \ } \left( {\rm tr} \left[ G(\omega + i\eta) \right] \right)
.
\end{eqnarray}


\subsection{Finite Temperature Average of Operators}

The Monte Carlo method for computing trace makes it possible to
evaluate the expectation value of an arbitrary operator $A$ at a finite
temperature $T$, which is defined as
\begin{eqnarray}
\label{eq:average}
\langle A \rangle_T &=& Z^{-1} {\rm tr}\left[ e^{-\beta H } A \right] 
\nonumber \\
&=&Z^{-1} \sum_n \langle n| e^{-\beta H } A |n \rangle \\
Z &=& {\rm tr}\left[ e^{-\beta H } \right]=\sum_n \langle n| e^{-\beta H } |n \rangle
\end{eqnarray}
where $\beta=1/T$ (we use $k_B=1$) and the sum runs over a chosen basis set of complete orthonormal basis $|n\rangle$. 
In principle, we can compute (\ref{eq:average}) by using the Monte Carlo scheme (\ref{eq:trace.monte}) for evaluating the trace.
However, the difficulty in evaluating the exponential operator
$e^{-\beta H }$ may hinder us from applying this straightforward scheme. To overcome this obstacle, we transform the expression by using the eigenkets of the Hamiltonian as a basis set, namely,
\begin{eqnarray}
\lefteqn{
 \langle A \rangle_T \times Z 
}
\nonumber \\
&=& {\rm tr}\left[ e^{-\beta H } A \right]  
=  \sum_\lambda e^{-\beta E_\lambda} \langle \lambda | A | \lambda \rangle \\
&=&  \sum_\lambda \int dE e^{-\beta E} \langle \lambda | \delta(H-E) A | \lambda \rangle \\
&=&  \sum_\lambda \int dE e^{-\beta E}
\nonumber \\
&& \times 
\langle \lambda | \frac{-1}{2i\pi} (G(E+i\eta) -G(E-i\eta)) A | \lambda \rangle \\
&=&  \frac{-1}{2i\pi} \int dE e^{-\beta E} 
\nonumber \\
&&\times {\rm tr}\left[ (G(E+i\eta)-G(E-i\eta)) A \right] \\
\label{eq:thermal.expect}
&=&  \frac{-1}{2i\pi} \int dE e^{-\beta E} 
\nonumber \\
&&\!\!\!\!\! \times \left( {\rm tr}\left[ A G(E+i\eta) \right]- {\rm tr} \left[ A^\dagger G(E+i\eta)\right]^* \right) 
.
\end{eqnarray}
If $A$ is a Hermitian operator, (\ref{eq:thermal.expect}) reduces to
\begin{eqnarray}
\label{eq:thermal.expect.hermite}
\lefteqn{
 \langle A \rangle_T \times Z 
}
\nonumber \\
&=&  \frac{-1}{\pi} \int dE e^{-\beta E} {\rm Im \ } \left( {\rm tr}\left[ A G(E+i\eta) \right] \right) 
.
\end{eqnarray}
The partition function $Z$ can be evaluated by using the unit matrix $I$ in place of $A$. 
Note that the imaginary part of the Green's functions works as a energy
filter function extracting the component of energy $E$ from the
random ket $| \Phi \rangle$ when we evaluate the trace.

\section{Numerical Examples}
\label{sec:examples}
%
%
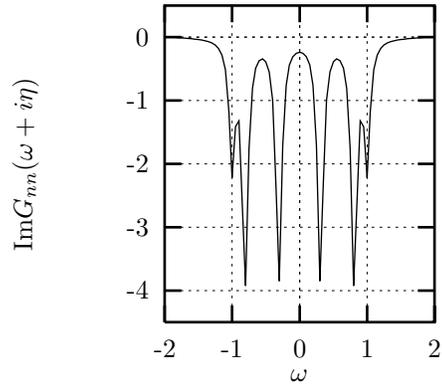
\begin{figure}[tbp]%
\begin{center} 
\setlength{\unitlength}{0.1bp}
\begin{picture}(1800,1696)(0,0)
\put(1108,51){\makebox(0,0){$\omega$}}
\put(100,848){%
\makebox(0,0)[b]{\shortstack{${\rm Im}G_{nn}(\omega+i\eta)$}}%
}
\put(1917,151){\makebox(0,0){.}}
\put(1617,151){\makebox(0,0){2}}
\put(1363,151){\makebox(0,0){1}}
\put(1109,151){\makebox(0,0){0}}
\put(854,151){\makebox(0,0){-1}}
\put(600,151){\makebox(0,0){-2}}
\put(540,1526){\makebox(0,0)[r]{.}}
\put(540,1326){\makebox(0,0)[r]{0}}
\put(540,1087){\makebox(0,0)[r]{-1}}
\put(540,848){\makebox(0,0)[r]{-2}}
\put(540,609){\makebox(0,0)[r]{-3}}
\put(540,370){\makebox(0,0)[r]{-4}}
\end{picture}
\end{center} 
\caption{${\rm Im}G_{nn}(\omega+i \eta)$ for $N=10$ and $\eta=0.1$.}
\label{fig:1}
\end{figure}

In this section, we show several numerical results to demonstrate the
effectiveness of Particle Source Method.
For simplicity, we calculate only one-body problems. However, these results include the Hamiltonian matrices of $N=10^6$, which is comparable to the dimension of the Hamiltonian matrices in manybody problems. Therefore we believe that  our method is the effective also in manybody problems. 
All numerical results in this section have been calculated with complex double precision arithmetic of FORTRAN.

\subsection{Perfect 1D Lattice}
Let us study the Hamiltonian of an electron in one 
dimensional space,
\begin{equation}
\label{eq:H}
H=\frac{p^2}{2 m_e} + V(x),
\end{equation}
where $m_e$ is mass of electron and $V(x)$ is the static potential.
After discretizing in space with the lattice size $\Delta x$, the Hamiltonian is approximated 
by a tight binding form,
\begin{eqnarray}
\label{eq:H.tight}
H &=& \frac{-\hbar^2}{2m_e\Delta x^2} 
\sum_{n=1}^{N}\left( c_n^\dagger c_{n+1} + c_n c_{n+1}^\dagger\right)
\nonumber \\
&& +\sum_{n=1}^{N} \left( \epsilon_n + \frac{\hbar^2}{m_e\Delta x^2} \right) c_n^\dagger c_n,
\end{eqnarray}
where $\epsilon_n=V(x_n)$ and $c_n^\dagger$ and $c_n$ are 
the creation and annihilation operator of electron at the site 
$x_n=n\times\Delta x$$\ (n=0,1,\cdots,N)$.
The periodic boundary condition is set as
\begin{equation}
\label{eq:boundary}
\langle n=0| \phi \rangle = \langle n=N | \phi \rangle
\end{equation}
where $|n\rangle$ is the electron state at the n-th site.

%
%
\begin{figure*}[ht]%
\begin{center}
\setlength{\unitlength}{0.1bp}
\begin{picture}(1800,1496)(0,0)
\put(1254,1182){\makebox(0,0)[r]{exact}}
\put(1254,1282){\makebox(0,0)[r]{PSM}}
\put(1108,51){\makebox(0,0){$\omega$}}
\put(100,848){%
\makebox(0,0)[b]{\shortstack{${\rm Re \ }G_{nn}(\omega+i\eta)$}}%
}
\put(1815,151){\makebox(0,0){.}}
\put(1515,151){\makebox(0,0){2}}
\put(1312,151){\makebox(0,0){1}}
\put(1109,151){\makebox(0,0){0}}
\put(905,151){\makebox(0,0){-1}}
\put(702,151){\makebox(0,0){-2}}
\put(540,1626){\makebox(0,0)[r]{.}}
\put(540,1326){\makebox(0,0)[r]{2}}
\put(540,1087){\makebox(0,0)[r]{1}}
\put(540,848){\makebox(0,0)[r]{0}}
\put(540,609){\makebox(0,0)[r]{-1}}
\put(540,370){\makebox(0,0)[r]{-2}}
\end{picture}
\setlength{\unitlength}{0.1bp}
\begin{picture}(1800,1496)(0,0)
\put(1254,1182){\makebox(0,0)[r]{exact}}
\put(1254,1282){\makebox(0,0)[r]{PSM}}
\put(1108,51){\makebox(0,0){$\omega$}}
\put(100,848){%
\makebox(0,0)[b]{\shortstack{${\rm Im \ }G_{nn}(\omega+i\eta)$}}%
}
\put(1815,151){\makebox(0,0){.}}
\put(1515,151){\makebox(0,0){2}}
\put(1312,151){\makebox(0,0){1}}
\put(1109,151){\makebox(0,0){0}}
\put(905,151){\makebox(0,0){-1}}
\put(702,151){\makebox(0,0){-2}}
\put(540,1626){\makebox(0,0)[r]{.}}
\put(540,1326){\makebox(0,0)[r]{2}}
\put(540,1087){\makebox(0,0)[r]{1}}
\put(540,848){\makebox(0,0)[r]{0}}
\put(540,609){\makebox(0,0)[r]{-1}}
\put(540,370){\makebox(0,0)[r]{-2}}
\end{picture}
\end{center} 
\caption{(a) real part and (b) imaginary part of $G_{nn}(\omega+i \eta)$ for $N=10^6$ and $\eta=10^{-3}$.}
\label{fig:1d}
\end{figure*}
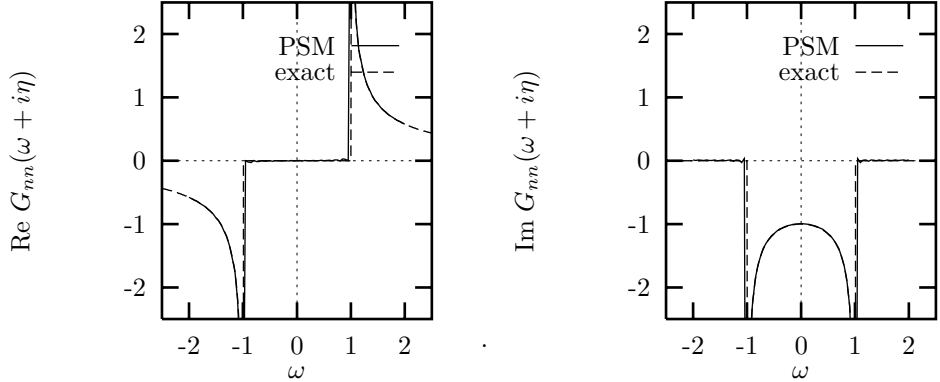

When $V(x)=0$, the exact analytical eigenstates and eigenvalues of the Hamiltonian (\ref{eq:H.tight}) with the 
boundary condition (\ref{eq:boundary}) are well known,
\begin{eqnarray}
\label{eq:exact.ket}
|E_m\rangle &=& A \sum_{n=1}^{N} \exp \left( 
i k_m n \Delta x \right) |i\rangle \\
\label{eq:exact.energy}
E_m &=&  \frac{\hbar^2}{m_e\Delta x^2}\left[1 - \cos\left(  k_m \Delta
x \right) \right] \\
k_m &=& \frac{m \pi}{N \Delta x}
\end{eqnarray}
where $A$ is a normalizing constant and $m$ is an integer 
$m=0,\pm 1, \pm 2,\cdots, \pm (N-2)/2, N/2$ for even $N$ and 
$m=0,\pm 1, \pm 2,\cdots, \pm (N-1)/2 $ for odd $N$. 
Note that (\ref{eq:exact.energy}) approximates well the parabolic dispersion relation (\ref{eq:H}) of the continuum model, if $m \ll N$ or $E \ll 1$.
In the following, we set $\hbar=m_e=\Delta x=1$ for simplicity.

Figure~\ref{fig:1} shows the imaginary part of the Green's function $G(\omega+i \eta)$ for $N=10$ and $\eta=0.1$, where $\omega=E-1$ is the energy measured from the band center. The numerical result reproduces faithfully the exact spectrum (\ref{eq:exact.energy}) of the Hamiltonian (\ref{eq:H}).

Figure~\ref{fig:1d} compares the Green's function $G_{nn}(\omega+i \eta)$ of a long perfect lattice calculated by the multiple frequency method to the exact analytical result.
For the numerical calculation, we used parameters, $\alpha=0.1$, $\eta=10^{-3}$, $\delta=10^{-2}$, and $\Delta \omega=5\times 10^{-2} $ and $N=10^6$ .
The computational time was 3 hours on the supercomputer at RIKEN.
The exact result in the limit $N\rightarrow \infty$ and $\eta \rightarrow +0$ is calculated by using the analytical expression$^{1)}$,
\begin{equation}
\label{eq:exact}
G_{nn}(\omega + i\eta) =
\left\{
\begin{array}{ll}
\displaystyle
\frac{-i}{\sqrt{1-\omega^2}} & (|\omega|<1) \\ 
\displaystyle
\frac{{\rm sgn}(\omega) }{\sqrt{\omega^2-1}} & (|\omega|>1)
\end{array}
\right.
.
\end{equation}

\subsection{Resonant Scattering by a square well potential}

The transmission probability of a particle described by the Hamiltonian (\ref{eq:H}) with an attractive rectangular potential,
\begin{equation}
V(x)= \left\{
\begin{array}{ll}
-V_0 & \mbox{for \ } |x-x_0|<a  \ \ \ (V_0 > 0) \\
0 & \mbox{for \ } |x-x_0|>a
\end{array}
\right.
\end{equation}
has an analytical expression$^{20)}$,
\begin{equation}
\label{eq:transmit.analytic}
T=
\frac{1}{
\displaystyle
1+\frac{V_0^2}{4E(E+V_0)} \sin^2\left( 2 a \sqrt{2(E+V_0)} \right)
}.
\end{equation}

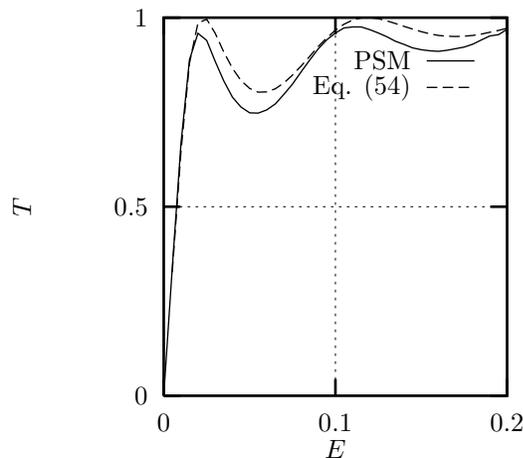
\begin{figure}[tb]%
\begin{center}
\setlength{\unitlength}{0.1bp}
\begin{picture}(2077,1728)(0,0)
\put(1531,1414){\makebox(0,0)[r]{Eq. (54)}}
\put(1531,1514){\makebox(0,0)[r]{PSM}}
\put(1247,51){\makebox(0,0){$E$}}
\put(100,964){%
\makebox(0,0)[b]{\shortstack{$T$}}%
}
\put(1894,151){\makebox(0,0){0.2}}
\put(1247,151){\makebox(0,0){0.1}}
\put(600,151){\makebox(0,0){0}}
\put(540,1677){\makebox(0,0)[r]{1}}
\put(540,964){\makebox(0,0)[r]{0.5}}
\put(540,251){\makebox(0,0)[r]{0}}
\end{picture}

\end{center}
\caption{Transmission probability $T_{LR}=\left| G_{LR}(\omega+i \eta)
v \right|^2$ as a function of energy. The parameters are $N=10^6$ and $\eta=10^{-4}$.}
\label{fig:3}
\end{figure}

Figure~\ref{fig:3} compares the transmission probability
$T_{LR}=\left| G_{LR}(\omega+i \eta)v \right|^2$ calculated by using 
PSM with the analytical result (\ref{eq:transmit.analytic}).
For the numerical calculation, we used parameters, $\alpha=0.1$, $\eta=10^{-4}$, $\delta=10^{-2}$, and $\Delta \omega=5\times 10^{-2} $ and $N=10^6$ .
The transmission probability calculated by PSM is slightly smaller
than the exact result. This is probably because of the finite imaginary part of the
energy, $\eta=10^{-4}$, which physically corresponds to the absorbtion 
of the particle.
The computational time for this calculation was 3 hours on the supercomputer at RIKEN.


\section{Summary}
\label{sec:summary}
In this article, we developed the PSM for calculating the
real time Green's functions of large sparse $N \times N$ Hamiltonian matrices, which exploits the numerical solution of the inhomogeneous time-dependent Schr\"{o}dinger equation.
The method has a clear-cut structure reflecting the most naive
definition of the Green's functions, and is very suitable to parallel and 
vector supercomputers. It requires, as the Lanczos method, memory
space of oder of $N$, and the CPU time of oder of $N$ for a given set
of parameters.
The PSM can also calculate matrix elements of the products of several Green's functions
and other operator, while the Lanczos method can calculate matrix
elements of operators that contains only one Green's function.
This is because PSM can calculate $N$ matrix elements,
$G(\omega+i\eta)|j\rangle$, at one calculation, while the Lanczos
method can calculate only one diagonal matrix element $\langle j|
G(\omega+i\eta)|j\rangle$ at a time.
We applied the PSM to simple lattice models and demonstrated that 
the method can be a powerful tool to study dynamical properties of
finite quantum systems.

%
%

\section*{Acknowledgment}
The author would like to acknowledge valuable discussions with A.~Mitsuktake, Y.~Okamoto, and M.~Suzuki.
The numerical calculations in this article were performed on NEC SX-3/34R at the Institute for Molecular Science (IMS), and on Fujitsu VPP500 at RIKEN.
This work was supported by Special Postdoctoral Researchers Program of RIKEN.

\vspace{\baselineskip}


References
\re
1) E.N.~Economou, {\it Green's Functions in Quantum Physics}, (Springer-Verlag, New York, 1983) ISBN 0-387-12266-4.
\re
2) P.A.~Lee and D.S.~Fisher, Phys. Rev. Lett. {\bf 47}, 882 (1981); D.J.~Thouless and S.~Kirkpatrick, J.~Phys. {\bf C14}, 235 (1981); A.~MacKinnon, Z.~Phys. {\bf B59}, 385 (1985).
\re
3) H.U.~Baranger, D.P.~DiVincenzo, R.A.~Jalabert, and A.D~Stone, Phys. Rev. {\bf B44}, 10637 (1991); T.~Ando, Phys. Rev. {\bf B44}, 8017 (1991).
\re
4) M.~Suzuki, S.~Miyashita, and A.~Kuroda, Prog. Theor. Phys. {\bf 58} 1377 (1977).
\re
5) For a review see, e.g., {\it Quantum Monte Carlo Methods}, edited by M.~Suzuki, (Springer, Berlin, 1987); W.~Linden, Phys. Rep. {\bf 220}, 53 (1992); E.Y.~Loh and J.E.~Gubernatis, in {\it Electronic Phase Transitions}, edited by W.~Hanke and Yu.~V.~Kopaev, (Elsevier, Amsterdam, 1992), p.177.
\re
6) R.N.~Silver, J.E.~Gubernatis, and D.S.~Sivia, Phys. Rev. Lett. {\bf 65}, 496 (1990).
\re
7) C.~Lanczos, J.~Res. Nat. Bur. Stand. {\bf 45}, 255 (1950); {\bf 49}, 33 (1952).
\re
8) For a review see, e.g., D.W.~Bullet, R.~Haydock, V.~Heine, and M.J.~Kelly, in {\it Solid State Physics} edited by H.~Erhenreich, F.~Seitz, and D.~Turnbull (Academic, New York, 1980), Vol. 35;  E.~Dagotto, Rev. Mod. Phys. {\bf 66}, 763 (1994).
\re
9) A.~Cordelli, G.~Grosso, G.P.~Parravicini, Comp. Phys. Comm. {\bf 83} 255 (1995).
\re
10) J.~Jaklic and P.~Prelovsek, Phys. Rev. {\bf B49} 5065 (1994).
\re
11) M.L.~Williams, and H.J.~Maris, Phys. Rev. {\bf B31}, 4508 (1985).
\re
12) J.~Schwinger, {\it Particles and Sources}, (Gordon \& Breach, 1969); L.H.~Ryder, {\it Quantum Field Theory}, (Cambridge University Press, Campridge, 1985) ISBN0-521-33859-X. 
\re
13) K.~Yakubo, T.~Nakayama, and H.J.~Maris, J.~Phys. Soc. Jpn. 
{\bf 60}, 3249 (1991); T.~Terao, K.~Yakubo, and T.~Nakayama, Phys. Rev. {\bf E50}, 566 (1994); T.~Nakayama, in {\it Computational Physics as a New Frontier in Condensed Matter Research} edited by H. Takayama, M. Tsukada, H. Shiba, F. Yonezawa, M. Imada and Y. Okabe, (Physical Society of Japan, Tokyo, 1995) ISBN4-89027-004-3.
\re
14) K.~Fukamachi, and H.~Nishimori, Phys. Rev. {\bf B49}, 651 (1994).
\re
15) A.~Askar and A.S.~Cakmak, J. Chem. Phys. {\bf 68}, 2794 (1978).
\re
16) C.~Leforestier, R.H.~Bisseling, C.~Cerjan, 
M.D.~Feit, R.~Friesner,
A.~Guldberg, A.~Hammerich, G.~Jolicard, W.~Karrlein, H.-
D.~Meyer, N.~Lipkin, O.~Roncero, and R.~Kosloff, J. Comp. Phys. {\bf 94}, 59 (1991), and references therein.
\re
17) T. Iitaka, Phys. Rev. E{\bf 49}, (1994) 4684;
T. Iitaka, N. Carjan, and D. Strottman, Comp. Phys. Comm. {\bf 90}, 251 (1995); T. Iitaka, {\it Introduction to Computational Quantum Dynamics}, (Maruzen, Tokyo, 1994) ISBN4-621-03971-7, (in Japanese). 
\re
18) M.D.~Feit, J.A.~Fleck, and A.~Steiger, J.~Comp. Phys. {\bf 47}, 412 (1982).
\re
19) S.T.~Kuroda, T.~Suzuki, Jpn. J. Appl. Math. {\bf 7}, 231 (1990).
\re
20) J.J.~Sakurai, {\it Modern Quantum Mechanics}, (Addison-Wesley,
Massachusetts, 1994) ISBN 0-201-53929-2.
\re
21) Y.~Okamoto and H.J.~Maris, Comp. Phys. Comm. {\bf 76}, 191 (1993);
A.~Mitsutake, T.~Iitaka, and Y.~Okamoto, Comp. Phys. Comm. (in press).

%
%

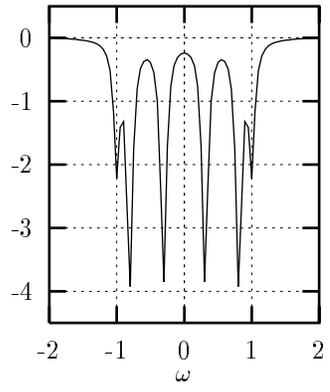
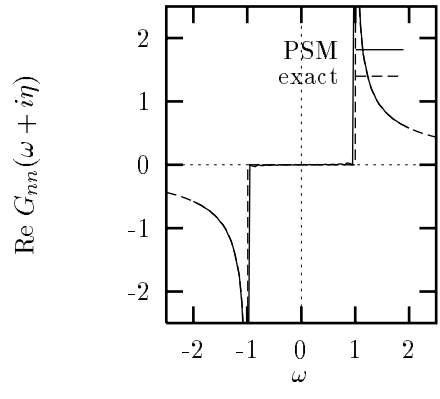
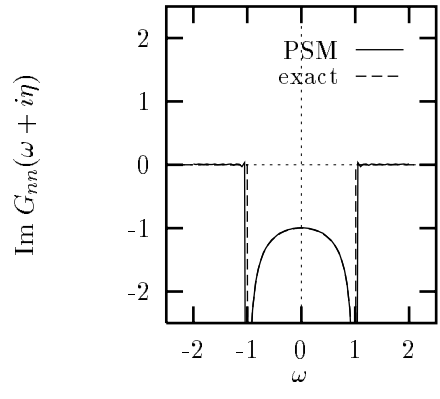
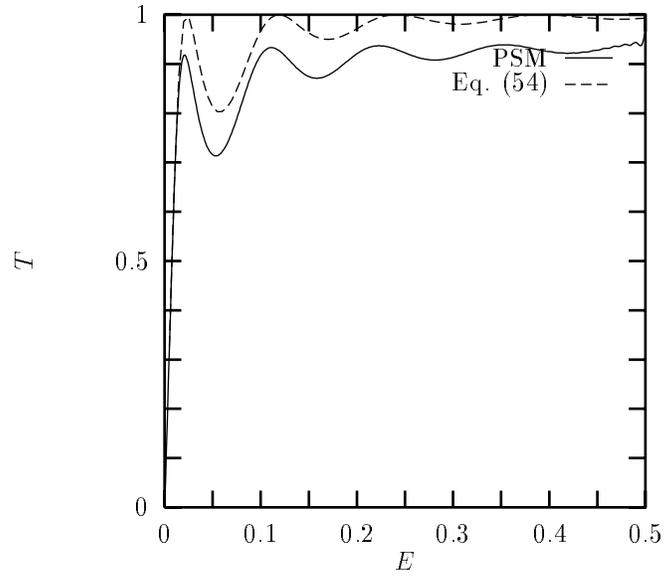
\end{document}